\documentclass[aps,pra,twocolumn,showpacs]{revtex4}
\usepackage{graphicx}
\usepackage{amsmath}
\usepackage{amssymb}

\newcommand{\ket}[1]{\ensuremath{\left|{#1}\right\rangle}}

\newcommand{\cnot}{\textsc{cnot}}
\newcommand{\oper}[1]{\boldsymbol{\mathsf{#1}}}

\begin{document}

\title{Hyperentanglement-assisted Bell-state analysis}

\author{S. P. Walborn}
\email[]{swalborn@fisica.ufmg.br}

\affiliation{Universidade Federal de Minas Gerais, Caixa Postal 702, 
Belo Horizonte, MG
30123-970, Brazil}

\author{S. P\'adua}

\affiliation{Universidade Federal de Minas Gerais, Caixa Postal 702, 
Belo Horizonte, MG
30123-970, Brazil}

\author{C. H. Monken}

\affiliation{Universidade Federal de Minas Gerais, Caixa Postal 702, 
Belo Horizonte, MG
30123-970, Brazil}

\date{\today}

\begin{abstract}
We propose a simple scheme for complete Bell-state measurement of 
photons using hyperentangled states - entangled in multiple degrees 
of freedom.  In addition to hyperentanglement, our scheme requires 
only linear optics and single photon detectors, and is realizable 
with current technology.  At the cost of additional classical 
communication, our Bell-state measurement can be implemented 
non-locally.  We discuss the possible application of these results to 
quantum dense coding and quantum teleportation. 
\end{abstract}


\maketitle

\section{Introduction}
Bell-state measurement (BSM) - distinguishing between the four 
maximally-entangled Bell states -  
is required in many quantum information schemes, including quantum 
dense coding 
\cite{bennett92a,mattle96}, quantum teleportation 
\cite{bennett93,bouwmeester97,boschi98} and entanglement swapping 
\cite{bennett93,pan98,jennewein02}.  However, 
it has been proven that a complete 
BSM (distinguishing between the four states with 100\% efficiency) is 
impossible using only linear operations and classical communication
\cite{vaidman99,lutkenhaus99,calsamiglia01,ghosh01}.  In fact, Ghosh 
\textit{et. al.}  \cite{ghosh01} have proven that, if only a single 
copy is provided, the best one can do is discriminate between two 
Bell states.  Calsamiglia and L\"utkenhaus \cite{calsamiglia01} have 
shown that the maximum efficiency for a linear Bell-state analyzer is 
$50\%$.   
\par
A favorable characteristic of the photon as a carrier of quantum 
information is the relative ease with which entangled photons can be 
created and transported.  Two-photon Bell states are easily generated 
using spontaneous parametric down-conversion (SPDC) in one of several 
degrees of freedom \cite{rarity90b,tapster94,kwiat95,kwiat99}.   
There are several methods for optical Bell-state measurement that 
allow one to  distinguish two 
\cite{weinfurter94,braunstein95,mattle96,michler96,walborn03b} of the 
four Bell-states states (resulting in 3 classes of states).  All of 
these methods use local or non-local two-photon interference effects 
at beam splitters.  For example, Mattle \textit{et. al.} 
\cite{mattle96} have performed an experimental demonstration of dense 
coding, in which the Bell-states were created in the polarization 
degrees of freedom of the two photons.  When the photons meet at a 
common beam splitter, the overall bosonic symmetry of the two-photon 
state requires that photons in the antisymmetric singlet state exit 
in different output ports, while the symmetric triplet states end up 
in the same output port \cite{zeilinger94}.  Polarization analyzers 
are then used to further discriminate among the triplet states.  
Weinfurter has proposed a BSM method using momentum-entangled photons 
that allows one to distinguish all four Bell-states with $25\%$ 
efficiency \cite{weinfurter94}.  It is possible to distinguish among 
the four Bell states using 
nonlinear optical processes  \cite{paris00,kim00a}, however, with 
present technology these methods suffer from low efficiency.  There 
have also been several proposals using two-photon absorption 
\cite{scully99, delre00}.  
\par
In recent years, some attention has been paid to Bell-state analysis 
using hyperentangled states \cite{kwiat98a,walborn03b}.  Utilizing 
entanglement in additional auxiliary degrees of freedom, it is 
possible to perform a complete BSM.  Due to the enlarged Hilbert 
space, this type of complete BSM is not restricted to the efficiency 
limits presented in 
\cite{vaidman99,lutkenhaus99,calsamiglia01,ghosh01}.  Kwiat and 
Weinfurter  \cite{kwiat98a} have proposed a scheme using photons 
entangled in polarization and momentum (spatial mode).  Their method, 
which relies on linear optics and two-photon interference effects,  
requires detectors that distinguish between one- and two-photon 
detection events.  We showed that this requirement on the detectors 
could be removed if the hyperentangled photons were created by SPDC 
using a Hermite-Gaussian pump beam \cite{walborn03b}.  The 
Hermite-Gaussian beam used is an odd function of the horizontal 
transverse spatial coordinate, which inverts the two-photon 
interference behavior \cite{walborn03a}, allowing for identification 
of all four Bell-states in coincidence detections.         
\par
Here we present a new method for complete Bell-state analysis using 
hyperentangled states.  This scheme differs from others in that it 
(\textit{i}) does not rely on two-photon interference, (\textit{ii}) 
does not require detectors sensitive to photon number, and 
(\textit{iii}) can be implemented non-locally with 2 bits of 
additional classical communication.  In section \ref{sec:hyper} we 
briefly discuss the creation of hyperentangled states.  We present 
our hyperentangled Bell-state analyzer in section \ref{sec:hbsa}, and 
discuss the application of these results to quantum information 
protocol such as dense coding and teleportation.         
\section{Hyperentanglement}
\label{sec:hyper}

 \begin{figure}
 \includegraphics[width=7cm]{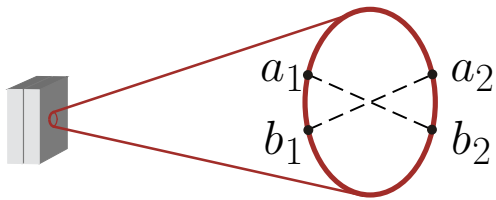}
 \caption{\label{fig:source} Hyperentangled states can be generated 
using the type-I two-crystal source \cite{kwiat99}.  Selecting two 
sets of regions ($a_{1}b_{2}$ and $a_{2}b_{1}$) gives photon pairs 
entangled in momentum and polarization.}
 \end{figure}

We will work with hyperentangled two-photon states of the form 
$\ket{\Pi} \otimes \ket{\eta} \equiv \ket{\Pi}\ket{\eta} $.  Here 
$\ket{\Pi}$ and $\ket{\eta}$ are the four dimensional vectors 
representing the polarization and momentum degrees of freedom of the 
two photons, respectively.  In the basis defined by linear horizontal 
($H$) and linear vertical ($V$) polarization, the 
polarization-entangled Bell-states are:
\begin{subequations}
\label{eq:1}
\begin{align}
\ket{\Psi^{\pm}} & = \frac{1}{\sqrt{2}}(\ket{H}_{1}\ket{V}_{2} \pm 
\ket{V}_{1}\ket{H}_{2} )  \\
\ket{\Phi^{\pm}} & = \frac{1}{\sqrt{2}}(\ket{H}_{1}\ket{H}_{2} \pm 
\ket{V}_{1}\ket{V}_{2} )
\end{align}
\end{subequations}
where $\ket{\sigma}_{j}$ stand for the polarization state of the 
photon $j$.  Likewise, the momentum-entangled Bell-states are:
\begin{subequations}
\label{eq:2}
\begin{align}
\ket{\psi^{\pm}} & = \frac{1}{\sqrt{2}}(\ket{a}_{1}\ket{b}_{2} \pm 
\ket{b}_{1}\ket{a}_{2} )  \\
\ket{\phi^{\pm}} & = \frac{1}{\sqrt{2}}(\ket{a}_{1}\ket{a}_{2} \pm 
\ket{b}_{1}\ket{b}_{2} )
\end{align}
\end{subequations}
where $\ket{a}_{j}$ and $\ket{b}_{j}$ represent different spatial 
modes of photon $j$.  We note here that polarization states have been 
denoted with uppercase letters and momentum states with lowercase 
letters.    
\par
For our hyperentangled-Bell-state analysis we will consider states of 
the form $\ket{\Pi}\ket{\psi^{+}}$, where $\ket{\Pi}$ is one of the 
polarization Bell states (\ref{eq:1}), though one could use any of 
the states (\ref{eq:2}) with similar results.  Recently, it was shown 
that this type of hyperentangled two-photon state could be used to 
violate a generalized form of the Greenberger-Horne-Zeilinger theorem 
\cite{chen03}, and may be useful in creating decoherence-free 
subspaces \cite{kwiat00}.  These states can be generated by means of 
spontaneous parametric down-conversion (SPDC) in several ways 
\cite{kwiat97,kwiat99,chen03}.  For example, the type-I two-crystal 
source \cite{kwiat99} emits polarization-entangled photons of the 
same wavelength around the rim of a cone (FIG. \ref{fig:source}).  In 
this source, crystal 1 emits pairs of (say) $H$-polarized photons and 
crystal 2 emits pairs of $V$-polarized photons in superimposed 
emission cones.  Phase-matching conditions guarantee that photon 
pairs are emitted on opposite sides of the cone.  If the two-crystal 
interaction region lies entirely within a coherence volume of the 
pump laser beam, a polarization- and momentum-entangled state of  the 
form $\ket{\Pi}\ket{\eta}$ can be selected by the two sets of regions 
$a_{1}b_{2}$ and $a_{2}b_{1}$.     
One can adjust the phase so that the momentum state is 
$\ket{\psi^{+}}$.  Half- and quarter-wave plates can be used to 
switch between the four polarization Bell-states \cite{kwiat95}.  

\section{Bell-state analysis}
\label{sec:hbsa}

 \begin{figure}
 \includegraphics[width=7cm]{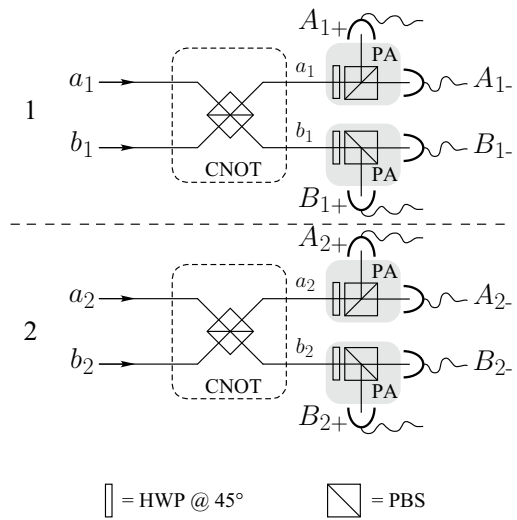}
 \caption{\label{fig:BSA_hyp}The hyperentangled Bell-state analyzer.  
The PBS are polarizing beam splitters in the $H-V$ basis.  HWP are 
half-wave plates.  The \cnot\ operation uses the polarization as the 
control and momentum as the target.  The PA are polarization 
analyzers in the $\pm 45^{\circ}$ basis (denoted as $\pm$) consisting 
of a HWP and a PBS.}
 \end{figure}

The hyperentangled-Bell-state analyzer is shown in FIG. 
\ref{fig:BSA_hyp}.  The hyperentangled state 
$\ket{\Pi}\ket{\psi^{+}}$ is first incident on a set of 
polarizing-beam splitters (PBS), which reflect $H$-polarized photons 
and transmit $V$-polarized photons.  The PBS perform a controlled-not 
(\cnot) logic operation between the polarization (control) and 
spatial (target) degrees of freedom \cite{simon02}.  If the 
polarization is $V$ then the photon is transmitted, and switches 
modes.  If the polarization is $H$, then the photon is reflected, and 
remains in the same mode.  The complete PBS operation on the 
polarization and spatial mode of photon $j$ is
\begin{align}
\ket{H}_{j}\ket{a}_{j} & \longrightarrow \ket{H}_{j}\ket{a}_{j} 
\nonumber \\
\ket{H}_{j}\ket{b}_{j} & \longrightarrow \ket{H}_{j}\ket{b}_{j} 
\nonumber \\
\ket{V}_{j}\ket{a}_{j} & \longrightarrow \ket{V}_{j}\ket{b}_{j} 
\nonumber \\
\ket{V}_{j}\ket{b}_{j} & \longrightarrow \ket{V}_{j}\ket{a}_{j}
\label{eq:3}
\end{align}  
It is straightforward to show that the four states 
$\ket{\Pi}\ket{\psi^{+}}$ transform as
\begin{align}
\ket{\Psi^{\pm}}\ket{\psi^{+}}& \longrightarrow  
\ket{\Psi^{\pm}}\ket{\phi^{+}}   \nonumber \\
\ket{\Phi^{\pm}}\ket{\psi^{+}}& \longrightarrow  
\ket{\Phi^{\pm}}\ket{\psi^{+}}   \nonumber \\
\label{eq:4}
\end{align}
A quick look at (\ref{eq:4}) shows that the PBS's mark the momentum 
state by performing a polarization-controlled logic operation.  The 
momentum state can then be used to discriminate between the 
$\ket{\Psi^{\pm}}$ and  $\ket{\Phi^{\pm}}$ polarization states:  
coincidence detections in modes $a_{1}a_{2}$ or $b_{1}b_{2}$ imply 
states $\ket{\Psi^{\pm}}$ while coincidences in $a_{1}b_{2}$ or 
$b_{1}a_{2}$ imply states $\ket{\Phi^{\pm}}$.    Using additional 
polarization analyzers (PA in FIG.  \ref{fig:BSA_hyp}) orientated at 
$45^{\circ}$, we can discriminate between the respective $\pm$ 
states.  Specifically, $\ket{\Psi^{+}}$ and $\ket{\Phi^{+}}$ give 
coincidence counts at the $+45,+45$ or $-45,-45$ output ports while 
$\ket{\Psi^{-}}$ and $\ket{\Phi^{-}}$ give coincidence counts at the 
$-45,+45$ or $+45,-45$ output ports.  A summary of the detector 
signatures for the polarization Bell states are shown in table 
\ref{tab:1}.  
\par
We note that all four Bell-states are recognized in the coincidence 
basis, with no need for detectors that are sensitive to photon 
number, and without the use of two-photon interference effects.    
Furthermore,  the BSM can be performed non-locally at the cost of 2 
bits of classical communication,   since the physicist measuring 
system $1$ needs to communicate one of four possible outcomes to 
system $2$ (or vice versa).

 \begin{table}
 \caption{\label{tab:1} Detector signatures for polarization 
Bell-states using the momentum state $\ket{\psi^{+}}$ as an ancilla.  
``$+$" $ \equiv +45^{\circ}$ and ``$-$"$ \equiv -45^{\circ}$.}
 \begin{ruledtabular}
 \begin{tabular}{cc}
state  & detector signature \\
$ \ket{\Psi^{+}}$ & $A_{1+}A_{2+}$ or $B_{1+}B_{2+}$ or 
$A_{1-}A_{2-}$ or $B_{1-}B_{2-}$ \\
$ \ket{\Psi^{-}}$ & $A_{1+}A_{2-}$ or $B_{1+}B_{2-}$ or 
$A_{1-}A_{2+}$ or $B_{1-}B_{2+}$ \\
$ \ket{\Phi^{+}}$ & $A_{1+}B_{2+}$ or $B_{1+}A_{2+}$ or 
$A_{1-}B_{2-}$ or $B_{1-}A_{2-}$ \\
$ \ket{\Phi^{-}}$ & $A_{1+}B_{2-}$ or $B_{1+}A_{2-}$ or 
$A_{1-}B_{2+}$ or $B_{1-}A_{2+}$ \\
\end{tabular}
 \end{ruledtabular}
 \end{table}
\par
The hyperentangled Bell-state analyzer is well within the scope of 
present technology, necessitating only wave plates, polarizing beam 
splitters and single photon detectors. 
\par
 A short analysis of the stability of the scheme proposed here may be 
in order.  In addition to mode overlap at the polarizing beam 
splitters, the use of the ancillary momentum state $\ket{\psi^{+}}$ 
requires phase stability between the modes.  A phase error $\alpha$  
of the form
\begin{align}
\ket{\psi^{+}} \longrightarrow & 
\frac{1}{\sqrt{2}}(\ket{a}_{1}\ket{b}_{2} + e^{i \alpha} 
\ket{b}_{1}\ket{a}_{2} ) \nonumber \\ 
= & \frac{1}{2}(1+e^{i \alpha})\ket{\psi^{+}} +  \frac{1}{2}(1 - e^{i 
\alpha})\ket{\psi^{-}} 
\label{eq:5}
\end{align}  
introduces an error in the BSM, since the \cnot\ operations with the 
ancillary momentum state $\ket{\psi^{-}}$ are:
\begin{align}
\ket{\Psi^{\pm}}\ket{\psi^{-}}& \longrightarrow  
\ket{\Psi^{\mp}}\ket{\phi^{-}},   \nonumber \\
\ket{\Phi^{\pm}}\ket{\psi^{-}}& \longrightarrow  
\ket{\Phi^{\mp}}\ket{\psi^{-}}.  
\label{eq:6}
\end{align}
The momentum states continue to show the same type of correlation, 
(the polarization states $\ket{\Psi^{\pm}}$ at $a_{1}a_{2}$ or 
$b_{1}b_{2}$, etc.), however, the polarization states have been 
switched.  From (\ref{eq:5}), the probability to obtain the correct 
Bell state is thus $|1+\exp(i\alpha)|^2/2$.  To avoid bilateral phase 
errors, we could use the momentum state $\ket{\psi^{-}}$ instead of 
the  $\ket{\psi^{+}}$ as an ancilla.  It has been shown that 
$\ket{\psi^{-}}$ is insensitive to collective decoherence 
\cite{kwiat00}.    
\par
\begin{figure}
 \includegraphics[width=7cm]{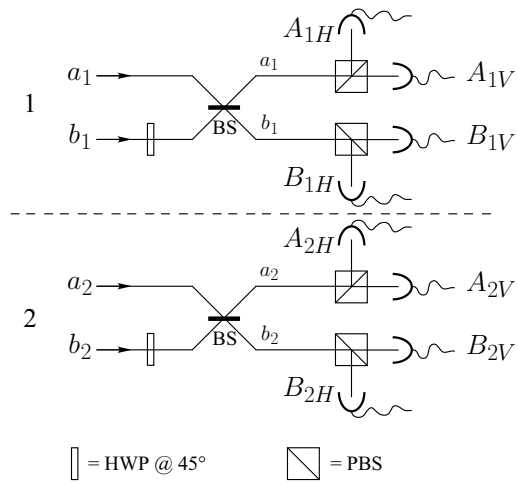}
 \caption{\label{fig:BSA_hyp_mom}The hyperentangled Bell-state 
analyzer for momentum Bell-states, using the polarization state as an 
ancilla.  The \cnot\ operation is now performed by the half-wave 
plates (HWP).  The BS are non-polarizing beam splitters.  The PBS are 
polarizing beam splitters in the $H-V$ basis.}
 \end{figure}

It is also possible to  use the polarization degrees of freedom as 
the ancilla and encode information in the momentum Bell-states.  A 
hyperentangled Bell-state analyzer for such an implementation is 
shown in FIG. \ref{fig:BSA_hyp_mom}.  The \cnot\ operation is 
performed  by half-wave plates (HWP) aligned at $45^\circ$ in modes 
$b_{1}$ and $b_{2}$.  The beam splitters (BS) and polarizing beam 
splitters (PBS) separate the 4 hyperentangled states.  The beam 
splitters transform the input modes as $a\longrightarrow (a + 
b)/\sqrt{2}$ and $b\longrightarrow (a - b)/\sqrt{2}$, which gives 
different correlations for the $\pm$ states.  The polarizing beam 
splitters separate the polarization ancilla states $\ket{\Phi^{+}}$ 
and $ \ket{\Psi^{+}}$ after the \cnot\ operation.  The detector 
signatures are shown in table \ref{tab:2}.  

 \begin{table}
 \caption{\label{tab:2} Detector signatures for momentum Bell-states 
using the polarization state $\ket{\Psi^{+}}$ as an ancilla.}
 \begin{ruledtabular}
 \begin{tabular}{cc}
state  & detector signature \\
$ \ket{\psi^{+}}$ & $A_{1H}A_{2H}$ or $B_{1H}B_{2H}$ or 
$A_{1V}A_{2V}$ or $B_{1V}B_{2V}$ \\
$ \ket{\psi^{-}}$ & $A_{1H}B_{2H}$ or $B_{1H}A_{2H}$ or 
$A_{1V}B_{2V}$ or $B_{1V}A_{2V}$ \\
$ \ket{\phi^{+}}$ & $A_{1H}A_{2V}$ or $B_{1H}B_{2V}$ or 
$A_{1V}A_{2H}$ or $B_{1V}B_{2H}$ \\
$ \ket{\phi^{-}}$ & $A_{1H}B_{2V}$ or $B_{1H}A_{2V}$ or 
$A_{1V}B_{2H}$ or $B_{1V}A_{2H}$ \\
\end{tabular}
 \end{ruledtabular}
 \end{table}

 \par

  We now discuss the possible application of hyperentangled 
Bell-state analysis to quantum information protocol.  The quantum 
dense coding protocol \cite{bennett92a} allows for the transmission 
of two bits of information in one quantum bit.  Two parties, Alice 
and Bob, each possess one photon of an entangled Bell-state 
(\ket{\Phi^{+}}, for example).  Since the reduced density matrix for 
Bob's photon is $\oper{I}/2$, where $\oper{I}$ is the $2 \times 2$ 
identity matrix, there is no information present in Bob's photon 
alone.  Some time later, Alice wishes to send a 2 bit message to 
Bob.  She switches among the four Bell-states using local operations, 
and then sends her photon to Bob, who performs a Bell-state 
measurement on the photon pair, and retrieves Alice's message.  Since 
there was no information present in Bob's photon, then the 2 bits of 
information present was sent in Alice's photon.  However, the reduced 
density matrix of Alice's photon is also $\oper{I}/2$, so an 
eavesdropper could not extract any information from Alice's photon 
alone.    
  \par 
  An appealing feature of the hyperentangled product states 
$\ket{\Pi}\ket{\eta}$ is that the two degrees of freedom can be 
manipulated independently.  For example, one can switch among the 
polarization Bell states using local operations (quarter- and 
half-wave plates in modes $a_{j}$ and $b_{j}$) on the polarization 
state, while leaving the momentum state untouched.  One could then 
implement a dense-coding scheme in which information is encoded in 
the polarization state, while the momentum state remains as an 
ancilla to assist in the complete Bell-state measurement.  We note 
that this implementation requires 4 qubits (encoded in 2 photons) to 
transmit 2 classical bits of information and thus it may be debatable 
as to whether this is actually ``dense" coding.  However, since the 
density matrix of (say) photon 2 is 
\begin{equation}
\hat{\rho}_{2} = \mathrm{trace}_{1}(\hat{\rho}_{12}) = 
\frac{1}{4}\oper{I}_{4},
\end{equation}
where $\hat{\rho}_{12}$ is the total $16 \times 16$ density matrix of 
photons 1 and 2 and $\oper{I}_{4}$ is the $4 \times 4$ identity 
matrix, we can still send 2 classical bits of information in one 
photon (containing  2 qubits) in such a way that an eavesdropper with 
access to only this photon cannot extract any information.
\par
Another use of a Bell-state measurement is in the quantum 
teleportation protocol \cite{bennett93}, which can be used to swap 
entanglement \cite{bennett93,pan98,jennewein02} and perform quantum 
logic operations for quantum computation \cite{gottesman99}.  In 
quantum teleportation, two spatially separate parties Alice and Bob 
each are in possession of one photon of an entangled pair, prepared 
in a Bell-state.  One of them, say Alice, wants to teleport the 
quantum state of a third photon to Bob.  Of course, Alice does not 
know what state she is teleporting, or else she could simply call Bob 
on the telephone and tell him to prepare that state.  Instead, Alice 
performs a Bell-state measurement on her two photons.  She then 
communicates classically to Bob, telling him the results of her BSM, 
at which point Bob, performs a local operation and recovers the state 
of the third photon, now encoded in his photon.  In this way Alice 
can teleport a quantum state to Bob without actually knowing what 
state she is sending.         
\par 
To date, quantum teleportation implementations \cite{bouwmeester97} 
can be performed with a Bell-state measurement that is 50\% efficient 
using two-photon interference, polarization analysis and single 
photon detectors.  Given an unknown quantum state, Alice, who shares 
a pair of maximally entangled photons with Bob, can teleport this 
state to him with 50\% efficiency.  Suppose Alice and Bob share a 
pair of polarization-entangled photons in the Bell-state 
$\ket{\Phi^{+}}_{12}$ and Alice would like to teleport the unknown 
polarization state $\ket{u}_{3}$ of photon 3.  To take advantage of 
hyperentangled Bell-state analysis, Alice would need to entangle the 
momentum degree of freedom of photon 2 with that of photon 3.  To do 
so requires a controlled logic operation between the two photons, 
such as a \cnot\ gate.  But if she could perform an efficient 
two-photon \cnot\ operation then she might as well use it to perform 
her BSM, which can be done with the same efficiency using a \cnot\ 
gate and a single photon Hadamard rotation \cite{chuang00}.
\par 
One might think that Alice could first entangle the momentum degrees 
of freedom of photons 2 and 3 using an inefficient \cnot\ gate, and 
discard the cases where the gate did not give the desired output 
(which is usually checked by measuring the ancillary modes).  She 
could then pass photon 3 on to a friend who performs some sort of 
complex quantum computation using the polarization as a qubit, and 
then passes it back to Alice for teleportation.  The complex quantum 
computation, which we assume to be much more difficult and time 
consuming than the teleportation protocol, need be performed only 
once, since the Bell-state measurement on photons 2 and 3 is now 
$100\%$ efficient.   However, since photon 3 is part of an entangled 
momentum state, it is now defined by 2 spatial modes.  So Alice's 
friend is required to run the computation on both spatial modes, 
which in principle is the same as running the computation twice, as 
is required on average in the $50\%$ efficient teleportation scheme.  
Thus,  a teleportation protocol using Bell-state measurement of 
hyperentangled states of this form does not present any gain over 
previous methods.    

\section{conclusion}

We have shown a simple method for complete Bell-state analysis using 
hyperentangled photons.  Our scheme requires only linear optics and 
single photon detectors, can be implemented non-locally with 2 bits 
of classical communication, and is well within the bounds of current 
technology.  We have briefly discussed the application of these 
results to the quantum teleportation protocol.  Given that our scheme 
requires photons entangled in multiple degrees of freedom, it does 
not provide any increase in efficiency to Bell-state measurements for 
quantum teleportation.  However, our method can be applied directly 
to implementations of quantum dense coding, resulting in the secure 
transmission of 2 bits of classical information in one photon 
(containing 2 qubits).

\begin{acknowledgments}
The authors acknowledge financial support from the Brazilian funding 
agencies CNPq, CAPES and Instituto do Mil\^enio de Informa\c{c}\~ao 
Qu\^antica.
\end{acknowledgments}

\end{document}